\documentstyle[aps,prd,amsmath,amssymb,amscd,tabularx,epsfig,preprint]{revtex}

\newcommand{\bra}[1]{\left\langle #1 \right|}
\newcommand{\ket}[1]{\left| #1 \right\rangle}

\begin{document}
\tightenlines
\title{The Bianchi Type $I$ minisuperspace model}
\author{Sigbj{\o}rn Hervik\footnote{e-mail:sigbjorn.hervik@fys.uio.no} \\
Department of Physics, University of Oslo\\
P.O.Box 1048 Blindern\\
N-0316 Oslo, Norway}
\maketitle

\begin{abstract}
The minisuperspace model of a Bianchi Type $I$ universe with compact 
spatial sections is investigated. The classical solutions are brought 
onto a form were the difference between compact and infinite spatial 
sections are manifest. One of the features of the compact case is that it 
has a non-trivial moduli space. The solution space of the compact Bianchi 
Type $I$ universe is 10 dimensional whereas the Kasner solutions only 
have a 1 dimensional solution space. We also include the classical 
solutions with dust 
and a cosmological constant. Solutions to the Wheeler-DeWitt equation 
are obtained in light of the tunneling boundary proposal by Vilenkin. 
Back reaction effects from a simple scalar field are also investigated.  
\end{abstract}

\section{Introduction}
Quantum cosmology was initiated by DeWitt \cite{dewitt} in the late 
sixties. The canonical quantization procedure which in the twenties was 
the key to the theory of Quantum mechanics, was again used to derive 
a ``Schr{\"o}dinger equation'' for gravity. The idea was that many of 
the cosmological mysteries which had their cause in the initial epoch of 
the universe, could be explained by  Quantum cosmology (QC). Through the 
work of DeWitt, Misner\cite{misner}, Ryan\cite{ryan72,ryan75,rs} etc., 
QC had prosperous days in the next decade. Suddenly the interest came to 
a hold, and the number of articles was drasticly reduced. But the 
interest for QC was again renewed in the 
eighties pioneered through the work of Hartle and Hawking\cite{hh}. They
 based their work 
on the boundary condition for a quantum state of the universe. Soon 
after, Vilenkin\cite{vil82,vil83,vil84,vil86,vil88,vil94,vv88} and Linde 
proposed their version of a suitable boundary condition.  

Most of the previous work on QC is based on FRW and deSitter universe 
models although some authors have studied anisotropic models (for 
instance \cite{louko}). The spatial sections of the FRW and deSitter 
models correspond to globally homogeneous, isotropic, simply connected 
spaces. In recent years 
a paper by the mathematician Thurston\cite{thurston} has drawn many a 
physicist's attention. His main conjecture, which still remains to be 
proven, claims that there are essentially 8 types of 3-manifolds. 
Concerning cosmology and especially QC the most interesting are their 
compact quotients which show interesting properties\cite{ll}.
In the nineties several authors investigated locally homogeneous 
spacetimes with non-trivial topologies. Hawking and Turok \cite{ht98} 
studied quantum creation of hyperbolic universes in the context of the no 
boundary proposal. Coule and Martin \cite{cm99} in addition to Costa
 and Fagundes \cite{cf98,cf99} studied cosmologies with compact hyperbolic
 spatial sections. We also want to mention the work of Fagundes 
\cite{fagundes82} already in the early eighties, in which he studies 
compact cosmologies which in the Thurston classification are quotients of 
${\mathbb R}\times {\mathbb H}^2$. In this paper we shall use a locally 
homogeneous anisotropic 
model that in the Thurston classification has the covering space $E^3$. 
We will not consider in any detail the covering space  $E^3$ itself 
(which is simply connected) but rather one of its compact quotients, the 
three torus, $T^3\cong {\mathbb R}^3/{\mathbb Z}^3$ (which is a 
multiply connected space with a non-trivial fundamental group). The 
possibility of multiply connected spatial sections of the Bianchi 
metrics \cite{fik,kodama} has again renewed the interest for QC. The 
space $T^3$ allows a Bianchi Type $I$ metric according to the Bianchi 
classification.

 Using Misner's notation the Bianchi metrics can be written as:
\[ ds^2=-N(t)^2dt^2+e^{2\alpha(t)}[e^{2\beta(t)}]_{ij}\chi^i\chi^j \]
where $\beta=\text{diag}(\beta_++\sqrt{3}\beta_-, \beta_+
-\sqrt{3}\beta_-, -2\beta_+)$ and $\chi^i=R_{ik}\omega^k$ where 
$R_{ik}$ is an $SO(3)$ matrix which can be parameterized by its 
\emph{Euler angles} $(\theta, \phi,\psi)$ i.e. ${\bf R}=e^{\phi 
\kappa_3}e^{\theta \kappa_1}e^{\psi \kappa_3}$. The matrices $(\kappa_1,
\kappa_2,\kappa_3)$ are the generators of the Lie group $SO(3)$. The 
invariant 1-forms $\omega^i$ obey the corresponding Bianchi Type Lie 
algebra: $d\omega^i=-\frac{1}{2}C^i_{jk}\omega^j\wedge\omega^k$. 

The Einstein-Hilbert action is given by\footnote{In this paper we use 
the following conventions: $c=16\pi G=\hbar=1$.} 
$S=\int_Md^4x\sqrt{-g}(R-2\Lambda)+2\int_{\partial M}d^3x\sqrt{h}K$. 
In the case of a spatial homogeneous spacetime like the Bianchi Types, 
an invariant basis of forms on the spatial hypersurfaces can be found. 
Thus, three of the dimensions in the action may be integrated. We assume 
that we have a topology of a 3-torus, $T^3$, and for simplicity's sake 
the volume will be set equal to 1. Choosing compact spatial sections are 
done for 
two reasons: Firstly because the spatial integration of the action is 
then finite, but also because the solution space of non-isomorphic 
metrics has a dimension equal to the dimension of the (true) phase 
space\cite{as}. 

\section{The moduli space of the toroidal Bianchi Type $I$ universe 
and the Principle of symmetric criticality}
Let us briefly review how we construct the torus $T^3$ as a quotient space
 of $E^3$. It is important that we  differenciate between the symmetry 
group that each hypersurface possesses and the symmetry group the 
hypersurfaces 
possess as spatial sections in the four dimensional manifold. For a more 
thorough investigation on this construction consult \cite{kodama}.

The full symmetry group of $E^3$ is $IO(3)$, translations and $O(3)$ 
rotations. The Bianchi metrics admit a 3 dimensional transitive group of 
isometries. The 3 dimensional group corresponding to the Bianchi Type $I$ 
Lie algebra is translations in three dimensions, thus isomorphic to 
${\mathbb R}^3$.  ${\mathbb R}^3$ is a normal Lie subgroup of $IO(3)$ 
acting simply transitive on the spaces ${\mathbb R}^3$. We will therefore 
consider the manifold ${\mathbb R}^3$ with a simply transitively symmetry 
group ${\mathbb R}^3$ defined in the obvious way. We will call this 
construction $\hat{\mathbb R}^3$. These spaces have exactly the symmetry 
allowed by the Bianchi Type $I$ Lie algebra.
To construct a three torus $T^3$ as a quotient of  $\hat{\mathbb R}^3$ we 
can do the following:
We find a freely and properly discontinuous subgroup $\Gamma$ 
of\footnote{The symmetry group of $\hat{\mathbb R}^3$ will actually be 
${\mathbb R}^3\tilde{\times}D_2$ where $D_2$ is the dihedral group and 
$\tilde{\times}$ is the semi-direct product. $D_2$ has 4 elements: The 
unit element and rotations around the 3 axis by an angle of $\pi$.} 
Sym$(\hat{\mathbb R}^3)$, and identify points in ${\mathbb R}^3$ under 
the action of $\Gamma$. This will ensure that the resulting quotient 
space is a smooth manifold\cite{thurston97}. We are interested in 
subgroups $\Gamma$ so that the resulting quotient has the topology of a 
three torus. These subgroups $\Gamma$ can be characterized by three 
vectors ${\bf a, b, c} \in {\mathbb R}^3$ so that $[{\bf a, b, c}]\in 
GL^+(3)$. These vectors will be the generators of the fundamental group of
 the torus $T^3$ and points in $\hat{\mathbb R}^3$ are identified under 
the action (which in this case is addition) of the vectors 
${\bf a, b, c}$. $\Gamma$ will then be $\Gamma=\{(n{\bf a}+m{\bf b}+k
{\bf c})\in {\mathbb R}^3|n,m,k\in{\mathbb Z}\}$. The moduli space can 
therefore be described by a matrix $[{\bf a, b, c}]\in GL^+(3)$ apart from
 the freedom of discrete modular transformations represented by 
conjugation with respect to matrices in $GL^+(3,{\mathbb Z})$. We will 
not fix the gauge with respect to these transformations in this paper. We 
will simply ignore them. The actual parameter space is therefore 
Teichm\"uller space. Through the homeomorphism 
$GL^+(3)\cong {\mathbb R}^6\times SO(3)$ we will parametrize the SO(3) 
sector by the Euler angles $(\theta,phi,\psi)$. 

In this paper we will follow a slightly different but equivalent point of 
view. We consider a cube $C$ in ${\mathbb R}^3$. We parametrize the cube 
by $C=\{(x,y,z)|0\leq x,y,z \leq 1\}$ and identify points on the boundary 
of the cube so that we have a toroidal topology. The moduli space will be 
determined by the linear mapping $l_{\bf L}({\bf x})={\bf L x}$. In the 
above description the matrix ${\bf L}$ is simply ${\bf L}=[{\bf a, b, c}]$
, ${\bf L}\in GL^+(3)$. 
For the Bianchi Type $I$ universe the invariant 1-forms can be written 
locally as $\omega^i=K^i_{~j}dx^j$ where $K^i_{~j}$ is a constant matrix. After 
identification the coordinates $x^i$  
become angular-like variables with periodicity 1. 

For a minisuperspace model one always has to ensure that the 
equations of motion obtained from a locally variational principle are 
equivalent to the Einstein field equations. If this is true we say that 
the {\it Principle of symmetric criticality} holds. It has been known for 
a while that the Bianchi Class B models fail to satisfy this principle
\cite{ryan74}. To prove consistency of the minisuperspace model used in 
this paper we will use a theorem by Torre and collaborators
\cite{torre98,aft99}. This theorem states 
\begin{quote}
The principle of symmetric criticality (PSC) is valid for for any metric field 
theory derivable from a local Lagrangian density if and only if the 
following 2 conditions are satisfied at each point $x$ in the region of 
spacetime under consideration.
\begin{enumerate}
\item[(1)] $H^q(G,I_x)\neq 0$
\item[(2)] $V_x^I\cap(V_x^I)_0=0$
\end{enumerate}
\end{quote}
We will explain the symbols as we verify the principle of symmetric 
criticality in our case.

$G$ is the symmetry group which is used for reduction. In our case we have
 $G={\mathbb R}^3\tilde{\times}D_2$ \cite{kodama}. The orbits of $G$ in 
the spacetime have dimension $q$, i.e. $q=3$. $I_x$ is the isotropy group 
for a point $x$. For a Bianchi Type $I$ universe with toroidal spatial 
sections each point $x$ has $I_x\cong {\mathbb Z}^3{\times}D_2\subset G$. 
$H^q(G,I_x)$ is the Lie algebra cohomology of $G$ relative to $I$. In our 
case it is easy to see that since $H^3(T^3)\neq 0$ where $H^3(T^3)$ 
is the deRahm cohomology class of the $T^3$, we will have 
$H^q(G,I_x)\neq 0$. The first condition is therefore fulfilled. 
$V_x$ is the vector space of symmetric rank 2 tensors at a point $x$ in 
the spacetime, and $V^I_x$ is the vector space of $I_x$-invariant 
symmetric rank 2 tensors at $x$. $(V_x^I)_0$ is the annihilator of $V^I_x$
, i.e. linear functions on $V$ which vanish on $V^I$. By for instance 
direct calculation it is not difficult by ordinary linear algebra to show 
that also the second condition holds. Alternatively we can use the fact 
that the group action acts {\em{transversally}}\cite{aft99,aft99b} on the 
bundle of metrics. 
Thus the principle of symmetric criticality holds for the Bianchi Type 
$I$ universe with toroidal spatial topology.

\section{The Kasner universe}
Kasner \cite{kasner} solved the Einstein field equations for a Bianchi 
Type $I$ universe with a vanishing energy-momentum tensor\footnote{Due 
to the fact that a cosmological constant may be viewed as a vacuum 
energy, we will treat a cosmological constant as a part of the 
energy-momentum tensor\cite{gron2}.} , $T^{\mu \nu}=0$ and spatial 
sections homeomorphic to the Euclidean three-space ${\mathbb R}^3$. 
The solution which now bears his name can be written as:
\begin{equation} \label{eq:Kasner}
ds^2=-dt^2+t^{\frac{2}{3}}\left[t^{\frac{4}{3}\cos(\gamma)}dX^2+
t^{\frac{4}{3}\cos(\gamma-\frac{2}{3}\pi)}dY^2+t^{\frac{4}{3}\cos(
\gamma+\frac{2}{3}\pi)}dZ^2\right]
\end{equation}
where $\gamma$ is an angular variable. If
 the three (complex) roots of the cubic
\begin{equation}\label{cubic}
 z^3-\frac{64}{27}e^{3i\gamma}=0 
\end{equation}
are denoted by $p_1$, $p_2$ and $p_3$, the real parts of $p_1,~p_2$ and 
$p_3$ are equal to the three exponents inside the square brackets in eq. 
\ref{eq:Kasner}. Due to eq. \ref{cubic} the following will also be true:
\begin{equation} \label{rooteq} 
p_1+p_2+p_3=p_1^2+p_2^2+p_3^2=0 
\end{equation} 

The Kasner solutions are therefore a 1-parameter family of solutions 
parameterized by the angular variable $\gamma$. This solution space is 
therefore usually called the Kasner circle. 

Let us denote the 2-parameter family parameterization of the flat 
Euclidean 3-space with metric 
\begin{equation}\label{met:euclid}
d\sigma_{\gamma}^2(t)=t^{\frac{4}{3}\cos(\gamma)}dX^2+
t^{\frac{4}{3}\cos(\gamma-\frac{2}{3}\pi)}dY^2+
t^{\frac{4}{3}\cos(\gamma+\frac{2}{3}\pi)}dZ^2 
\end{equation}
by $({\mathbb R}^3,d\sigma^2_{\gamma}(t))$. For a fixed $X,~Y$ and $Z$ 
this family will be volume preserving and homogeneous but it has an 
anisotropic expansion (in $t$). 

Despite that Kasner solved the field equations for a vanishing source 
tensor, the solutions for a mixture of dust and a vacuum energy can be 
brought onto a similar form to that of Kasner eq. \ref{eq:Kasner}.
\section{The general solution of the Bianchi Type $I$ universe with 
$T^3$ spatial sections and with dust and a cosmological constant}
From \cite{kodama} we get the Hamiltonian
\begin{equation}\label{hamiltonian}
{\mathcal H}=\frac{N}{24}\left[e^{-3\alpha}(-p_{\alpha}^2+p_+^2+p_-^2)
+48M+48\Lambda e^{3\alpha}\right]
\end{equation}
where $M$ is a constant defined by $\int d^3x\sqrt{h}\rho_{dust}=2M$.
The canonical 1-form is given by:
\[\Theta=p_{\alpha}d\alpha+p_+d\beta_++p_-d\beta_-+p_{\theta}d\theta+
p_{\phi}d\phi+p_{\psi}d\psi\]
We have here treated the dust field non-dynamically, i.e. solved the 
classical equations for co-moving dust and thereafter inserted the 
solutions into the action. The conjugate momentum for the dust field is 
therefore absent in eq. \ref{hamiltonian} (compare with \cite{kt,salopek})

The Einstein field equations are now equivalent to the set of 
Hamiltonian equations
\begin{equation}
-\dot{p}_i  =\frac{\partial \mathcal{H}}{\partial q_i}, \quad
\dot{q}_i  =\frac{\partial \mathcal{H}}{\partial p_i}
\end{equation}
together with the constraint equation
\[ {\mathcal H}=0 \]
First we will choose the gauge $N=1$. 
 From the Hamiltonian equations of motion the following variables will be 
constants:
The Euler angles $(\theta,\phi,\psi)$ and their conjugated momenta 
$(p_{\theta},p_{\phi},p_{\psi})$ 
and the conjugated momenta $p_{\pm}$. Let us set
\[ p_+=4a, \quad p_-=4b. \]
and introduce the anisotropy parameter $A^2=a^2+b^2$. In addition two 
more constants $C_{\pm}$ appear when the equations for 
$\dot{\beta}_{\pm}$ are integrated. The constants 
$(\theta,\phi,\psi,p_{\theta},p_{\phi},p_{\psi}, A, C_+, C_-)$ have a 
clear geometrical meaning. These constants (for $A\neq 0$) specify a 
matrix\footnote{For an explicit expression of this matrix consult 
\cite{kodama}.} ${\bf{L}}\in GL^+ (3)$ such that the regular solid cube 
is mapped into the spaces $({\mathbb R}^3,d\sigma^2_{\gamma}(t))$ by the 
linear mapping $l_{\bf L}(\bf{x})\equiv L{\bf{x}}$. The cube is mapped 
onto a parallelepiped with volume $A=\det({\bf L})$.   
Introducing the volume element $v=e^{3\alpha}$, the constraint equation 
yields the  equation for $v$:
\begin{equation}\label{eq:volum}
\dot{v}^2=3\Lambda v^2+3Mv+A^2
\end{equation}
The constant $M$ will scale as $A$, so it is more useful to introduce a 
new constant $m$ by $M=Am$. The volume element $v$ will also scale as 
$A$  so we introduce the function $V(t)$ by 
$v(t)=AV(t)$. 

\subsection{The case $\Lambda>0$}
A singularity seems unavoidable since for $v=0$, $\dot{v}=\pm A$, the 
volume element starts off at the initial singularity or collapses at the 
final singularity at a speed $A$. Our interest will be  mainly on the 
expanding solutions. Defining $\kappa=\frac{m}{2\Lambda}$ and 
$\omega^2=3\Lambda$ we can write the solutions as:
\[ V(t)=\frac{1}{\omega}[\sinh\omega t+\kappa \omega (\cosh\omega t -1)] 
\]
where we have chosen $V(0)=0$. 

The remaining equations can be integrated in a straight forward manner, 
and introducing
\[ \Sigma_+(t)=\frac{\frac{2}{\omega}(\cosh\omega t -1)}{\sinh\omega t+
\kappa \omega (\cosh\omega t -1)} \]
the equations for $\beta_{\pm}$ can be solved to yield
\begin{align}
\beta_+=&\frac{1}{3}\sin(\gamma-\frac{\pi}{6})\ln \Sigma_+(t)+C_+ \\
\beta_-=&\frac{1}{3}\cos(\gamma-\frac{\pi}{6})\ln \Sigma_+(t)+C_-
\end{align}
The line-element for a dust-filled Bianchi Type $I$ with a positive 
cosmological constant is therefore
\begin{equation}
ds^2=-dt^2+V(t)^{\frac{2}{3}}\left[
\Sigma_+(t)^{\frac{4}{3}\cos(\gamma)}dX^2
+\Sigma_+(t)^{\frac{4}{3}\cos(\gamma-\frac{2}{3}\pi)}dY^2
+\Sigma_+(t)^{\frac{4}{3}\cos(\gamma+\frac{2}{3}\pi)}dZ^2\right]
\end{equation} 
where\footnote{Here the notation is a bit misleading(which it often is 
concerning angular variables). The ``exact'' differentials $d{\bf X}$ 
and $d{\bf x}$ are only exact in the covering space ${\mathbb R}^3$. On 
the torus $d{\bf x}$ can be a globally defined 1-form which will be 
closed but not exact i.e. it corresponds to a non-trivial element in the 
deRahm cohomology class of the torus.} $d{\bf{X}} ={\bf{L}}d\bf{x}$ and 
$0\leq x, y, z\leq 1$ are ``angular'' variables. 

\subsection{The case $\Lambda<0$}
Let us now consider the case where $\Lambda$ is negative. Perhaps the 
simplest way to obtain the solution for $\Lambda<0$ is to redefine 
$\omega^2=3|\Lambda|$, and $\kappa=\frac{m}{2|\Lambda|}$, and perform 
the following mapping:
\begin{align*}
\omega & \longmapsto i\omega \\
\kappa & \longmapsto -\kappa
\end{align*}
to obtain for the volume function
\[ V(t)=\frac{1}{\omega}[\sin\omega t+\kappa \omega (1-\cos\omega t )]. \]
We can then define 
\[ \Sigma_-(t)=\frac{\frac{2}{\omega}(1-\cos\omega t) }{\sin\omega t
+\kappa \omega (1-\cos\omega t )} \]
The resulting line-element is then
\begin{equation}
ds^2=-dt^2+V(t)^{\frac{2}{3}}\left[
\Sigma_-(t)^{\frac{4}{3}\cos(\gamma)}dX^2
+\Sigma_-(t)^{\frac{4}{3}\cos(\gamma-\frac{2}{3}\pi)}dY^2
+\Sigma_-(t)^{\frac{4}{3}\cos(\gamma+\frac{2}{3}\pi)}dZ^2\right]
\end{equation}
This line element describes a universe that expands at first. After a 
while the cosmological term becomes dominant and turn the expanding 
phase into a contracting one, and it ends in a final singularity after 
an elapsed time 
$ t_F=\frac{2}{\omega}\left[\pi-\arctan\left(\frac{2}{m}
\sqrt{\frac{|\Lambda|}{3}}\right)\right] $.

\subsection{The case $\Lambda=0$, and a general form of the line element}
The equations for $\Lambda=0$ can be solved in a similar manner, and we 
will therefore just write down the result. Defining
\[ \Sigma_0(t)=\frac{t}{1+\frac{3}{4}mt} \]
the line element for $\Lambda=0$ takes the form
\begin{equation}
ds^2=-dt^2+\left(t+\frac{3}{4}mt^2\right)^{\frac{2}{3}}\left[
\Sigma_0(t)^{\frac{4}{3}\cos(\gamma)}dX^2
+\Sigma_0(t)^{\frac{4}{3}\cos(\gamma-\frac{2}{3}\pi)}dY^2
+\Sigma_0(t)^{\frac{4}{3}\cos(\gamma+\frac{2}{3}\pi)}dZ^2\right]
\end{equation}
Up to a simple rescaling these three line elements are those derived by 
Saunders\cite{saunders}. The $\Lambda=0$ case was also derived by 
Stephani\cite{stephani} and the $m=0$ case by Gr\o n\cite{gron}. 
Already now we can see the resemblance of these three cases and the 
Kasner line element. However, the line element can be brought into a 
more Kasner-like form by choosing another function $N$. 

\begin{figure}
\centering
\epsfig{figure=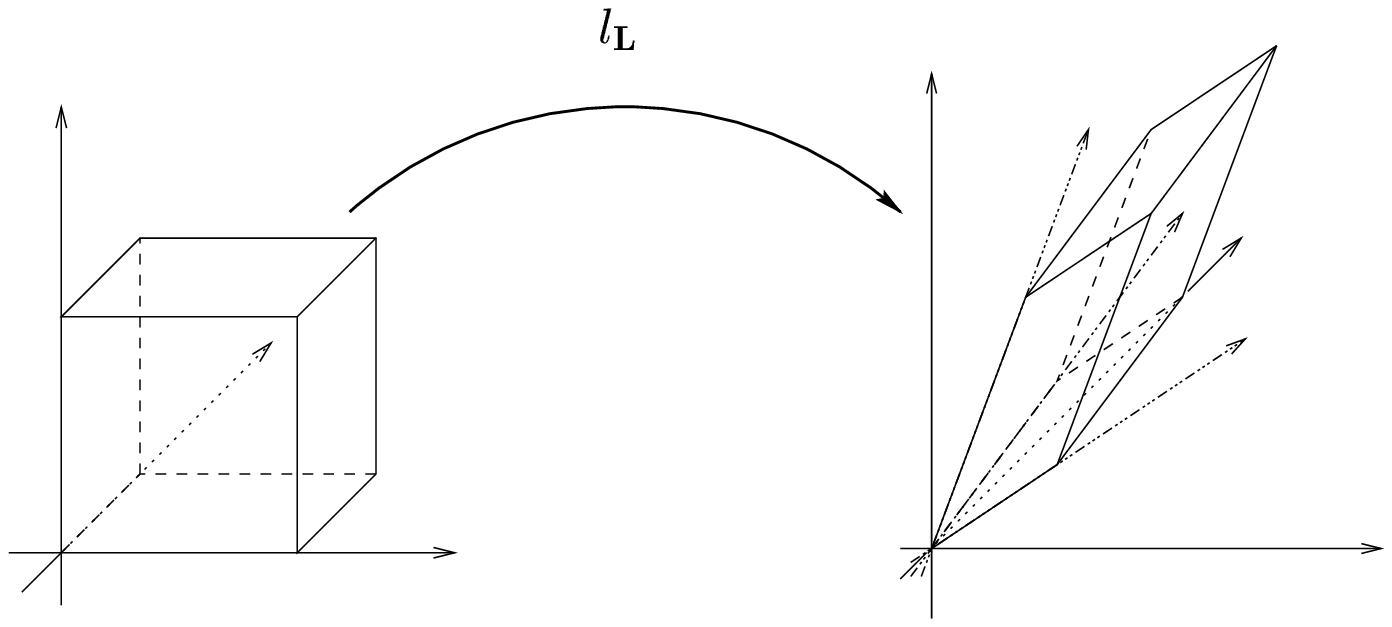, width=13cm}
\caption{Deformation of the cube: The mapping $l_{\bf L}:C\longmapsto ({\mathbb R}^3,d\sigma_{\gamma}^2(t))$}\label{GLmap}
\end{figure}

We define a new time variable, $\tau$, by:
\begin{eqnarray} 
t\equiv \int_0^{\tau}\frac{d z}{(1-\frac{3}{4}mz)^2-\frac{3}{4}\Lambda 
z^2}=\begin{cases}
\Sigma^{-1}_-(\tau) & ,\Lambda<0 \\
\Sigma^{-1}_0(\tau) & ,\Lambda=0 \\
\Sigma^{-1}_+(\tau) & ,\Lambda>0
\end{cases}
\end{eqnarray}
where $^{-1}$ means the inverse function.
Inserting this new time variable into the previous line elements, we see 
that all the three cases can be brought into a similar form. Thus we can 
write all the three line elements as:
\begin{align}\begin{split}\label{eq:genr}
ds^2=&-\frac{d\tau^2 }{\left((1-\frac{3}{4}m\tau)^2-\frac{3}{4}\Lambda 
\tau^2\right)^2} \\ &+\frac{1}{\left((1-\frac{3}{4}m\tau)^2-\frac{3}{4}
\Lambda \tau^2\right)^{\frac{2}{3}}}
\tau^{\frac{2}{3}}\left[\tau^{\frac{4}{3}\cos(\gamma)}dX^2+
\tau^{\frac{4}{3}\cos(\gamma-\frac{2\pi}{3})}dY^2+
\tau^{\frac{4}{3}\cos(\gamma+\frac{2\pi}{3})}dZ^2 \right]
\end{split}\end{align}
As we see, the line element is now manifestly cast onto a Kasner form. 
The solution space is now explicitly decomposed into the following 
sequence:
\begin{equation}\label{facseq}
\begin{CD}
C @>{\begin{CD}SO(3)\times{\mathbb R}^6 \\ @VVV \\ l_{\bf{L}}\end{CD}}>> 
{\mathbb{R}}^3 @>{\begin{CD}S^1 \\@VVV \\s_{\gamma}(t)\end{CD}}>>
({\mathbb R}^3,d\sigma^2_{\gamma}(t))
@>{\begin{CD} T^{\mu \nu} \\ @VVV \\ K\end{CD}}>> 
({\mathbb R}\times \Sigma, ds^2)
\end{CD}
\end{equation}
These mappings can be given the following interpretations:
\begin{enumerate}
\item[$l_{\bf{L}}$ :] Deformation of the cube. If we represent the unit 
cube $C$ by the embedding 
$C\cong\{(x^i), i=1,2,3|0\leq (x^i-x_0^i)\leq 1 \}$ for any $(x^i_0)$, 
then $l_{\bf{L}}({\bf x})=\bf Lx$ for ${\bf L}\in GL^+(3)\cong SO(3)
\times {\mathbb R}^6$.(See figure \ref{GLmap}) 
\item[$s_{\gamma}(t)$ :] An explicit 2-parameter parameterization of the 
Euclidean 3-space with an anisotropic metric. Given 
$(X,Y,Z) \in {\mathbb R}^3$ the metric is given by eq. \ref{met:euclid}.
\item[$K$ :] The Kasner foliation, given explicitly by: 
$K({\mathbb R}^3,d\sigma^2_{\gamma}(t))= ({\mathbb R}\times \Sigma, ds^2)$ 
where $\Sigma={\mathbb R}^3$ and 
$ds^2=-N(t)^2dt^2+V(t)^{\frac{2}{3}}d\sigma_{\gamma}^2(t)$. 
The $N$ and $V$ are determined by the Einstein field equations and as 
shown they are given by $N(t)=\frac{1}{\left((1-\frac{3}{4}mt)^2
-\frac{3}{4}\Lambda t^2\right)}$ and $V(t)=t\cdot N(t)$.
\end{enumerate}
The solution space is now parameterized by these mappings, and the 
effect of choosing compact spatial sections is now explicitly shown. 
We have also factorized the solution space into a topological 
(or modular) sector (the $SO(3)\times {\mathbb R}^6$ sector) and a 
Kasner sector (the Kasner circle). Contributions from the dust and the 
cosmological terms are also factored out. Different matter 
configurations only affect the last mapping $K$, all the other 
degrees of freedom affect only the mappings $l_{\bf{L}}$ and $s_{\gamma}$. 

Interestingly, the Kasner limit, where $A\longrightarrow \infty$ in such 
a way that $\lim_{A\longrightarrow \infty}l_{\bf{L}}(C)={\mathbb R}^3$, 
is well defined. The topological sector becomes degenerate and 
effectively we have only a $S^1$ degree of freedom. On the other hand
 the FRW limit 
$A\longrightarrow 0$ is apparently badly defined. Again the 
topological sector becomes degenerate in addition to the $S^1$-sector, 
but we do not seem to uniquely reproduce the FRW solutions. 

There is also one more interesting consequence of a non-trivial topology. 
For the Kasner universe ($T^{\mu\nu}=0$) the special case $\gamma=0$:
\[ds^2=-dt^2+t^2dX^2+dY^2+dZ^3\]
is through the transformation $\tilde{T}=t\cosh X,~\tilde{X}=t\sinh X$ 
seen to be equal to the inside of the future lightcone in the 
$(\tilde{T},\tilde{X})$ Minkowski space. Thus it can be naturally 
expanded to a flat manifold containing the apparent singularity $t=0$. 
The compact case on the contrary can not. Assuming 
${\bf L}=A^{\frac{1}{3}}\cdot({\bf id})$ the differential structure of 
the $(\tilde{T},\tilde{X})$ space is that of a cone. The singularity 
$t=0$ is now the locus of the cone. It therefore represents a true 
singularity in the $C^{\infty}$ structure of the maximal extended space 
(i.e. the cone including the locus). Interestingly we will actually know 
the true nature of the singularity. Taking the limit 
$A\longrightarrow \infty$ to obtain the Kasner universe, topological 
considerations suggest that the lightcone in the ``flat'' 
$(\tilde{T},\tilde{X})$ ought to be identified as \emph{one} point. The 
geometry of the space $\gamma=0$ is no longer globally flat, it is more 
like that of a cone. This explains perhaps why the ``flat space'' case 
$\gamma=0$ does not evolve as a flat space through the factorization 
\ref{facseq}.

\section{The Wheeler-DeWitt Equation}

In the previous sections we have solved the classical equations for a 
Bianchi Type $I$ universe. We will now quantizise the dynamical degrees of
 freedom according to the Wheeler-DeWitt procedure. The constraint 
equation will then turn into an operator equation on the wave function. 
The resulting equation is the so-called Wheeler-DeWitt (WD) equation. Let 
us also include a real scalar field $\Phi$, which later will yield 
interesting back-reaction effects. 

The WD equation for a Bianchi Type $I$ with a scalar field reads:
\[ \frac{1}{2}(-\nabla^2+V_E)\Psi=0 \]
where 
\begin{align*}
\nabla^2=& \frac{e^{-3\alpha}}{12}\left(
-e^{\xi \alpha}\frac{\partial}{\partial \alpha}e^{-\xi \alpha}
\frac{\partial}{\partial \alpha}+\nabla^2_{\beta}+
12\frac{\partial^2}{\partial \Phi^2}\right) \\
V_E=& V(\Phi)e^{3\alpha}+4M
\end{align*}
Here $\xi$ represents some of the factor ordering ambiguity and 
$\nabla^2_{\beta}=\frac{\partial^2}{\partial \beta_+^2}+
\frac{\partial^2}{\partial \beta_-^2}$ is the usual Laplace operator in 
$\beta$-space. 
To simplify the expressions we introduce the volume element 

$v=e^{3\alpha}$ and do the rescaling: $\tilde{\beta}_{\pm}=3\beta_{\pm}$, 
$\tilde{\Phi}=\frac{\sqrt{3}}{2}\Phi$, $\mu=\frac{16}{3}M$ and $\tilde{V}
(\tilde{\Phi})=
\frac{4}{3}V(\Phi)$. We now drop the tildes to avoid unnecessary writing. 
The WD equation then turns into (with $\zeta=1-\frac{\xi}{3}$)
\begin{equation}
\left[v^2\frac{\partial^2}{\partial v^2}+\zeta v 
\frac{\partial}{\partial v}-\nabla^2_{\beta}-\frac{\partial^2}{\partial 
\Phi^2}+\mu v+V(\Phi)v^2\right]\Psi(v,\beta_{\pm},\Phi, \phi, \theta, \psi
) =0
\end{equation}
Since the WD equation does not involve any of the Euler angles, the 
$SO(3)$ sector can be treated separately. 

\subsection{The $SO(3)$-sector}
Since $SO(3)\cong {\mathbb P}^3$, this section of superspace is naturally
equipped with an elliptic Riemannian structure. Due to the compactness of 
$SO(3)$ as a topological space there exists a complete countable set  of 
orthogonal functions on this space. Thus the conjugated momenta  are 
quantized. We may choose the eigenfunctions in such a way that they are 
characterized by three quantum numbers: $(l,m, m')$. These numbers are 
all integers and obeys: $ -l\leq m', m\leq l$ and $l=0,1,2..$. The 
eigenfunctions are not surprisingly the irreducible representation 
matrices of the group $SO(3)$, often written as 
$D^{(l)}_{mm'}(\phi, \theta, \psi)$. These functions are given by:
\begin{equation}
D^{(l)}_{mm'}(\phi, \theta, \psi)=e^{-im\phi}e^{-im'\psi}d^{(l)}_{mm'}
(\theta)
\end{equation}
where 
\begin{eqnarray} 
d^{(l)}_{mm'}(\theta)=\sum_{\lambda}\frac{(-1)^{\lambda}
[(l+m)!(l-m)!(l+m')!(l-m')!]^{\frac{1}{2}}}{\lambda !(l+m-\lambda)!
(l-m'-\lambda)!(\lambda+m'-m)!}\\
\times \left(\cos\frac{\theta}{2}\right)^{2l-2\lambda-m'+m}
\left(-\sin\frac{\theta}{2}\right)^{2\lambda+m'-m}
\end{eqnarray}
where the sum is only over integer $\lambda$ which makes sense. The 
algebraic properties of these functions now follow from the group 
theoretical properties of the Lie group $SO(3)$. We could for instance 
create a ``Trace wave function''. Every point in ${\mathbb P}^3$ 
(or element in $SO(3)$) may be represented by a unit rotation vector 
$\hat{n}$ in  ${\mathbb R}^3$ and a rotation angle $\alpha$. Then the 
``Trace wave function'' may be defined by:
\[
\Psi_{Tr}^{(l)}(\phi, \theta, \psi)=\sum_{m=-l}^l D^{(l)}_{mm}(\phi, 
\theta, \psi)=\frac{\sin(l+\frac{1}{2})\alpha}{\sin\frac{\alpha}{2}}
\]

\begin{figure}
\centering
\epsfig{figure=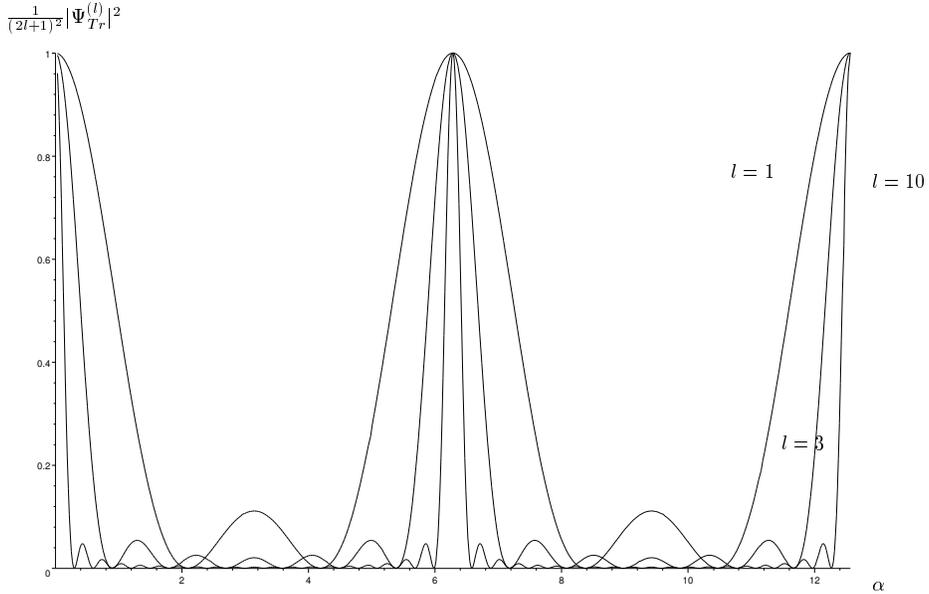, width=13cm}
\caption{The square of the Trace wavefunctions $\Psi^{(1)}_{Tr},
~\Psi^{(3)}_{Tr}$ and $\Psi^{(10)}_{Tr}$ normalized so that 
$\Psi^{(l)}_{Tr}=1$ at the unit element. Note also that the $x$-axis 
covers the range of $\alpha$ twice.}\label{tracesquared}
\end{figure}

These functions have their maximum at the unit element, will be zero 
where $\alpha=\frac{2\pi n}{2l+1}$ where $n$ is an integer 
$1\leq n \leq 2l$. From a classical quantum description if the Bianchi 
Type $I$ universe was in such a state, the fundamental cube  would look 
as if it were ``wobbling'' around its identity mapping. It might suddenly 
do a quantum leap to a different orientation. Investigating the 
``Probability''  amplitude $|\Psi_{Tr}^{(l)}|^2$ we see that the 
classical description describes a shell model (fig. \ref{tracesquared}). 
There are $2l$ regions where the Bianchi universe could be. These regions 
are separated by forbidden regions. The most probable shell is the one 
containing the identity element. Let us therefore call this the ground 
state. This situation becomes even more evident in the limit 
$l\longrightarrow\infty$. It might appear that these wave functions allow 
a geometry change. If a Bianchi Type $I$ universe was born in an 
``excited'' state which is meta-stable, it could suddenly cross the 
barrier to the ground state. The state in the neighborhood of the 
identity element would be the most probable. 
Due to the representation decomposition 
\[{\bf D}^{(l_1)}\otimes{\bf D}^{(l_2)}\cong
\bigoplus_{l=|l_1-l_2|}^{l_1+l_2}{\bf D}^{(l)}\]
the following will also be true:
\[
\Psi_{Tr}^{(l_1\otimes l_2)}=\Psi_{Tr}^{(l_1)}\cdot\Psi_{Tr}^{(l_2)}=
\sum_{l=|l_1-l_2|}^{l_1+l_2}\Psi_{Tr}^{(l)}
\] 
Thus the solution space spanned by these Trace wave functions could be 
generated by the two generators: $\Psi_{Tr}^{(0)}$ and $\Psi_{Tr}^{(1)}$. 
These Trace wave functions are of course only a particular set of 
functions which is by no means complete (although orthogonal) but drawn 
to attention only because of the algebraic beauty they possess. 
The famous orthogonality theorem for the matrix elements of irreducible 
representations will ensure that our functions 
$D^{(l)}_{mm'}(\phi, \theta, \psi)$ are orthogonal over the Riemannian 
space ${\mathbb P}^3$.
All of these identities follow from the Lie group properties of $SO(3)$ 
and there are of course a lot more than the ones mentioned here. Thus a 
lot of the properties of the Bianchi Type $I$ are directly related to the 
structure of the superspace, at least in this description. The fact that 
the superspace has this  Lie group structure provides a lot of 
information.

\subsection{Exact solutions for a zero-mass scalar field}
For a Klein-Gordon field the scalar potential can be written as 
$V(\Phi)=\lambda+m^2\Phi^2$, where $\lambda$ correspond to a cosmological 
constant. For this type of potential exact solutions for the WD equation 
is very difficult if not impossible to obtain. However, if we are dealing 
with a zero-mass scalar field $m=0$, exact solutions are possible to 
obtain. Firstly we can separate out the $\beta_{\pm}$ and $\Phi$ 
dependent parts by introducing $\Psi(v,\beta_{\pm},\Phi)=
F(v)e^{i(k_+\beta_++k_-\beta_-)}e^{in\Phi}$. As we see 
$k^2\equiv k_+^2+k_-^2$ corresponds to the classical anisotropy parameter 
$A^2$ i.e. the volume of the fundamental domain. Thus, these particular 
solutions are planar wave solutions where $F(v)$ satisfy the equation:
\begin{equation} \label{redWD} 
v^2\frac{d^2 F}{d v^2}+\zeta v \frac{dF}{d v}+(k^2+n^2+\mu v+\lambda v^2)F=0 
\end{equation}
This is \emph{Whittaker's equation} and its solutions can be 
written as
\begin{equation} 
F(v)=v^{a-\frac{1}{2}}W_{\mp L,p}(\pm 2i\lambda^{\frac{1}{2}}v) 
\end{equation}

where $a=\frac{\xi}{6}$, $L=\frac{i\mu}{2\lambda^{\frac{1}{2}}}$, 
$p^2=a^2-(k^2+n^2)$ and $W_{L,p}(z)$ is the 
Whittaker function. These function have an essential singularity at 
infinity and have a branch point at the origin.
This is due to the fact that the classical solutions have a physical 
singularity at vanishing 3-volume. In the absence of dust $\mu=0$, the 
equation \ref{redWD} turn into a Bessel equation. Thus the solution for 
$\mu=0$ can be written as 
\[ F(v)=v^aZ_p(\lambda^{\frac{1}{2}}v) \]
where $Z_p(z)$ is one of the Bessel functions or a linear combination of 
them. This can also be seen from the identity
\[ W_{0,p}(2z)=\sqrt{\frac{2z}{\pi}}K_p(z) \]
The behavior of these solutions are not drasticly altered with the 
inclusion of a non-zero dust term. For simplicity's sake we will set 
$\mu=0$ in the rest of this paper. In the further investigations the dust 
term is not essential and because the properties of the Bessel functions 
is better known we will simply drop the dust term. The 
solutions are therefore (a linear combination of) the Bessel functions.

In the case of a classical state that is not 
allowed, for instance if we have a negative cosmological constant and a 
large enough $v$, we would expect an exponential decrease of the 
wave function. For negative $\lambda$ we can write the solutions as a 
linear combination of the modified Bessel functions 
$K_p(|\lambda|^{\frac{1}{2}}v)$ and  $I_p(|\lambda|^{\frac{1}{2}}v)$. 
However,  $I_p(|\lambda|^{\frac{1}{2}}v)\approx 
e^{|\lambda|^{\frac{1}{2}}v}$ for  $|\lambda|^{\frac{1}{2}}v\gg |p|$, so 
we have to abandon these solutions if we demand an exponential decrease 
for  $|\lambda|^{\frac{1}{2}}v\gg |p|$. The actual linear combination is 
a matter of boundary condition, which we will leave for another section.

\subsection{A non-zero mass scalar field: Harmonic oscillator expansion}
For a non-zero mass scalar field the WD equation is difficult if not 
impossible to solve exactly. 
 
Let us try to find solutions of the form:
\begin{equation}\label{eq:harmexp}
\Psi(v, \Phi)=\sum_{n=0}^{\infty}c_n(v)\psi_n(v,\Phi)
\end{equation}
where $\psi_n(v,\Phi)$ are the eigenfunctions of the harmonic oscillator 
equation:
\begin{equation}\label{eq:harmosc}
\left[ -\frac{1}{2}\frac{\partial^2}{\partial \Phi^2}+\frac{1}{2}m^2v^2
\Phi^2\right]\psi_n=mv(n+\frac{1}{2})\psi_n
\end{equation}
Thus, we can consider the wave equation $\Psi$ as a path in the Hilbert 
space which is spanned by the basis $\{ \ket{n} \}$ where $\ket{n}$ is 
the usual normalized eigenstate of the one-dimensional harmonic 
oscillator equation.
In a coordinate representation these are given by:
\begin{equation}
\psi_n(v,\Phi)=\left(\frac{mv}{\pi}\right)^{\frac{1}{4}}
\frac{1}{\sqrt{2^n n!}}e^{-\frac{1}{2}mv\Phi^2}H_n(\Phi\sqrt{mv})
\end{equation}
where $H_n$ is the $n$'th Hermite polynomial. Calculating the derivatives 
with respect to $v$:
\begin{align}\label{eq:env}
v\frac{\partial}{\partial v}\ket{n}= & \left[\frac{1}{4}
-\frac{1}{2}mv\Phi^2\right]\ket{n}+\sqrt{\frac{n}{2}}\sqrt{mv}
\Phi\ket{n-1} \\ 
\begin{split} 
v^2\frac{\partial^2}{\partial v^2}\ket{n}= & \left[-\frac{3}{16}
-\frac{1}{4}mv\Phi^2+\frac{1}{4}m^2v^2\Phi^4\right]\ket{n} \\
& -\sqrt{\frac{n}{2}}(mv)^{\frac{3}{2}}\Phi^3\ket{n-1} \\
& +\frac{1}{2}\sqrt{n(n-1)}mv\Phi^2\ket{n-2}
\end{split} \label{eq:tov}
\end{align}
where we have used the identity:
\[H_n'=2nH_{n-1}\]
Using  the annihilation and creation operators $a$ and $a^{\dagger}$ 
defined by:
\begin{align}\label{eq:aop}
\begin{split}
a=& \sqrt{\frac{mv}{2}}\Phi-\frac{1}{\sqrt{2mv}}\frac{\partial}{\partial 
\Phi} \\
a^{\dagger}=& \sqrt{\frac{mv}{2}}\Phi+\frac{1}{\sqrt{2mv}}
\frac{\partial}{\partial \Phi}
\end{split} 
\end{align}
we can write any power of $\sqrt{mv}\Phi$ as
\[ (mv)^{\frac{k}{2}}\Phi^k=\frac{1}{\sqrt{2^k}}\left(a+a^{\dagger}\right)^k 
\]
Inserting eq. \ref{eq:harmexp} into the original WD equation for a 
non-zero mass scalar field, and using eq. \ref{eq:harmosc}, \ref{eq:env}, 
\ref{eq:tov} and \ref{eq:aop} we can get a differential equation which 
the coefficients $c_n(v)$ shall satisfy. If we instead introduce a new 
family of functions defined by $d_n(v) =v^{\frac{\zeta}{2}}c_n(v)$ the 
resulting equation may be written as:
\begin{align}
\begin{split}
0= &v^2d_n''+\left(\lambda v^2+mv(2n+1)+k^2+\frac{1}{8}-\frac{\xi^2}{36}
-\frac{1}{8}n(n+1)\right)d_n \\ 
&-\frac{1}{2}v\left(\sqrt{n(n-1)}d_{n-2}'-\sqrt{(n+1)(n+2)}d_{n+2}' 
\right) \\
&+\frac{1}{4}\left(\sqrt{n(n-1)}d_{n-2}-\sqrt{(n+1)(n+2)}d_{n+2} \right) 
\\
& +\frac{1}{16}\left(\sqrt{n(n-1)(n-2)(n-3)}d_{n-4}+\sqrt{(n+1)(n+2)(n+3)
(n+4)}d_{n+4} \right)
\end{split}
\end{align}
The infinite set of equations divides into two, one odd set of equations 
and one even set of equations, which involve respectively odd or even 
coefficients only. The Hilbert space is therefore divided into two: odd 
and even states. 
Now it is time for some approximations. Let us investigate the properties 
of the solutions in the limit $v\longrightarrow 0$. In the further 
calculations we will assume that the wave function $\Psi(v, \Phi)$ is 
independent of $\Phi$ in the limit $v\longrightarrow 0$ in the following 
sense:
\begin{quote}
Given a compact interval $I\subset {\mathbb R}$ and a $\delta>0$. Then 
there exists a $\epsilon>0$ so that for $0<v<\epsilon$, $|\frac{\partial 
\Psi}{\partial \Phi}(v, \Phi_0)|<\delta$ $\forall \Phi_0 \in I$.
\end{quote}
This is exactly the right boundary condition for  demanding that 
$c_n(v)\longrightarrow a_n(v)$ as $v\longrightarrow 0$ where $a_n(v)$ is 
the coefficients of a function independent of $\Phi$ (constant in $\Phi$) 
in the harmonic oscillator expansion. Using the test function 
$f=\left(\frac{mv}{4\pi}\right)^{\frac{1}{4}}$ we see that:
\[ \ket{f}=\sum_{n=0}^{\infty}\frac{\sqrt{(2n)!}}{2^nn!}\ket{2n} \]
Based on these results we can split the total wave function into an even 
and an odd part $\Psi=\Psi_{odd}+\Psi_{even}$.  We extract the solution 
for the constant function $f$ by defining a new family of coefficients by 
$d_{2n}=\frac{\sqrt{(2n)!}}{2^nn!}{\mathcal E}_n$. The equation for 
${\mathcal E}_n$ will then be
\begin{align} \label{eq:constexp} \begin{split}
0=& v^2{\mathcal E}_n''+\left(\lambda v^2+mv(4n+1)+k^2+\frac{1}{8}
-\frac{\xi^2}{36}-\frac{1}{2}n(n+\frac{1}{2})\right){\mathcal E}_n \\
& -v\left( n{\mathcal E}_{n-1}'-(n+\frac{1}{2}){\mathcal E}_{n+1}'\right) 
\\
&+\frac{1}{2}\left( n{\mathcal E}_{n-1}-(n+\frac{1}{2})
{\mathcal E}_{n+1}\right) \\
& +\frac{1}{4}\left(n(n-1){\mathcal E}_{n-2}+(n+\frac{1}{2})
(n+\frac{3}{2}){\mathcal E}_{n+2}\right)
\end{split} \end{align}
This equation can without difficulty be solved to first order in $v$. The 
result is:
\begin{equation}\label{eq:firstorder}
{\mathcal E}_n=v^l\left(1-\frac{1}{4}\frac{1}{1\pm\sqrt{\frac{\xi^2}{36}
-k^2}}(4n+1)mv+{\mathcal O}(v^2)\right)
\end{equation}
where $l=\frac{1}{4}\pm\sqrt{\frac{\xi^2}{36}-k^2}$. Assuming $\xi<6$ the 
real value of the first order term will be  negative, i.e. to first order 
the higher exited modes in the expansion will decay compared to the 
ground state. 

\subsection{Backreaction Effects: The massless case}
The preceding results describe an even wave function in $\Phi$. The 
expectation value of the scalar field vanishes, but the expectation value 
of $\Phi^2$ diverges as $v\longrightarrow 0$. This will have consequences 
for the effective Hamiltonian in the theory. Let us try to replace the 
energy momentum tensor, $T^{\mu \nu}$, by its expectation value 
$<T^{\mu \nu}>$ using the  obtained results. This effectively means that 
the scalar field operator is replaced with its expectation value:
\begin{align*}
-\frac{\partial^2}{\partial \Phi^2}+m^2v^2\Phi^2 \longmapsto 
&\frac{\bra{\Psi}\left[-\frac{\partial^2}{\partial \Phi^2}+m^2v^2\Phi^2
\right]\ket{\Psi}}{\left\langle \Psi|\Psi\right\rangle }
\end{align*}
In the massless case the expectation value of the energy momentum tensor 
only contributes with a positive constant $n^2$. Keeping the volume of 
the universe fixed (i.e. keeping $A^2$ fixed), the term $n^2$ will reduce 
the radius of the Kasner circle (the ``radius'' in eq. \ref{cubic}). Note 
also that the equations \ref{rooteq} will still be true. Since the center 
of the Kasner circle represents an isotropic universe, we see that the 
inclusion of a massless scalar field effectively reduces the anisotropy 
of the universe. 
\subsection{Backreaction Effects: The Harmonic oscillator expansion}
Using the harmonic oscillator expansion we can try to understand the 
effect of a non-zero mass scalar field on the classical equations. In the 
harmonic oscillator expansion we can write
\begin{align}
\bra{\Psi}\left[-\frac{\partial^2}{\partial \Phi^2}+m^2v^2\Phi^2\right]
\ket{\Psi}=mv\sum_{n=0}^{\infty}|c_n|^2(2n+1)
\end{align}
In our case we shall use only the even part, but unfortuneatly the 
results obtained can not be used as they stand; every sum will diverge. 
We have to regularize the result so that the sums may be performed.
Let us write:
\[ |{\mathcal E}_n|^2 \propto v^{2\text{Re}(l)} \left(1-\frac{2}{\eta}
\kappa mv\right)^{\eta (n+\frac{1}{4})} \] 
where $\kappa=\text{Re}\left(\frac{1}{1\pm\sqrt{\frac{\xi^2}{36}-k^2}}
\right)$ and $\eta$ is some unknown parameter which represents the 
uncertainty in the higher order contribution in the expansion 
(\ref{eq:firstorder}). To first order, however, we see that this is 
exactly the previously obtained results. With this replacement the sums 
may be performed and the result is to first order in $v$:
\begin{equation}
\frac{\bra{\Psi_{even}}\left[-\frac{\partial^2}{\partial \Phi^2}
+m^2v^2\Phi^2\right]\ket{\Psi_{even}}}{\left\langle \Psi_{even}
|\Psi_{even}\right\rangle}\approx mv\left(\frac{2\eta-1}{\eta}\right)
+\frac{1}{\kappa} 
\end{equation}
Luckily the unknown parameter $\eta$ does not contribute to lowest order, 
so the validity of the regularization procedure which we used is 
strengthened. 
If we assume $\xi=0$ and we associate $k^2$ with the anisotropy parameter 
$A^2$, we see that we get an effective anisotropy parameter 
$A^2_{eff}=2A^2+\frac{9}{16}$. Comparing this to the classical solutions, 
an increase of the effective anisotropy parameter in the Hamiltonian 
constraint correspond to an increase of the speed of the volume element 
by $\frac{A_{eff}}{A}$ and a decrease of the anisotropy speed by the same 
amount. 

Even though these considerations were very simple and only approximative, 
this indicates that the effect of a simple scalar field with mass is that 
 it reduces the effective anisotropy. Comparing with the zero mass 
solutions, we see that a zero mass scalar field also has the same effect, 
it increases the anisotropy parameter but reduces the anisotropy. The 
result in the limit $v\longrightarrow 0$ for a non-zero mass is thus 
reasonable.
 
\section{Boundary Conditions} 
Concerning boundary conditions, quantum cosmology has mostly invoked two 
different boundary conditions. One is that of Hartle-Hawking\cite{hh}, 
which is stated in terms of the path integral formalism. The 
Hartle-Hawking wave function $\Psi_{HH}$ for a Bianchi Type $I$ universe 
with a positive cosmological constant has been derived by Duncan and 
Jensen\cite{dj}. Duncan and Jensen also use three-torus as spatial 
sections, but they do not give any comments concerning the  $SO(3)$ 
sector. Apparently the only reason they use compact spatial sections is 
to make the spatial integration in the action finite. In this paper we 
will however be more interested in the tunneling boundary proposal by 
Vilenkin\cite{vil83}.

\subsection{The Bianchi Type $I$ minisuperspace}

Essential in this discussion is the description of the Bianchi Type $I$ 
from a minisuperspace point of view. 
If we exclude a scalar field degree of freedom, the minisuperspace will 
be 6-dimensional. Topologically we can write the minisuperspace as 
$M^3\times {\mathbb P}^3$ where ${\mathbb P}^3\cong SO(3)$ and $M^3$ is 
the 3-dimensional Minkowski space. As the $SO(3)$-sector has already been 
discussed, we will only look at the evolution in the Minkowski sector. We 
will also choose the lapse to be unity $N=1$. The DeWitt metric on the 
Minkowski sector will then be:
\begin{equation}\label{dewittmetric}
ds^2=e^{3\alpha}(-d\alpha^2+d\beta_+^2+d\beta_-^2)
\end{equation}

Doing a conformal transformation we can map this metric onto a 
Penrose-Carter diagram. 

\begin{figure}
\centering
\epsfig{figure=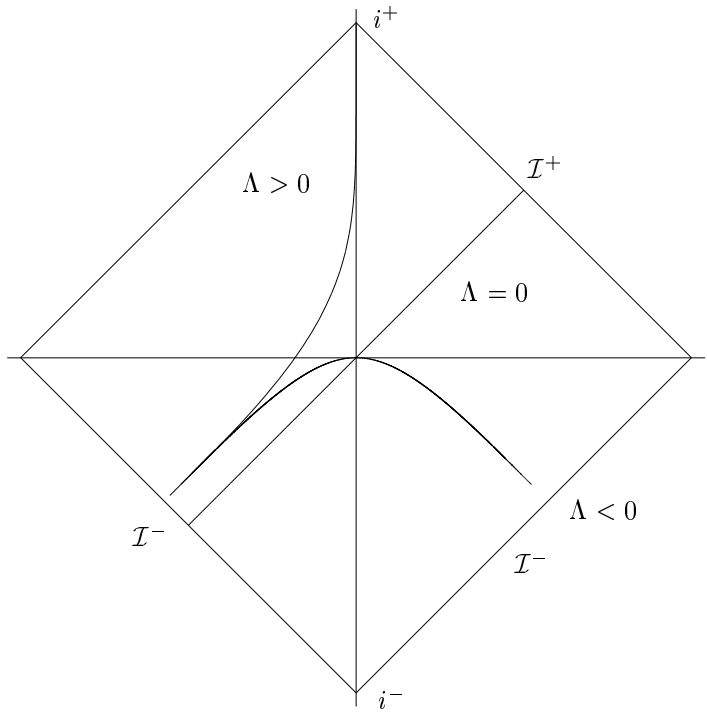,height=10cm, angle=0}
\caption{The evolution of the Bianchi Type $I$ in conformal superspace}
\label{biaenevol}
\end{figure}

\begin{figure}
\centering
\epsfig{figure=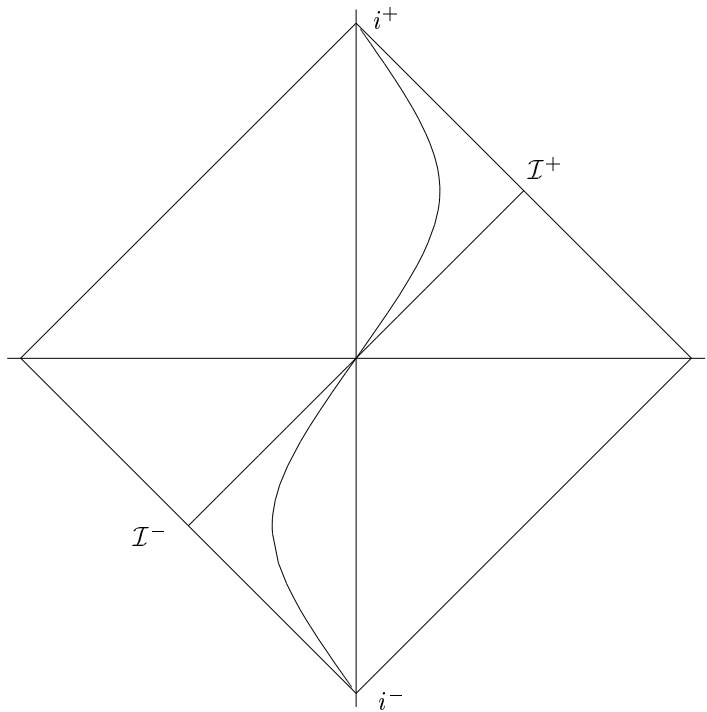, height=10cm}
\caption{The evolution of a Kasner universe versus the evolution of a 
Kasner universe with a reduced ``speed''}\label{timelikekasner}
\end{figure}

Mapping the classical solutions (with no dust) into this conformal 
diagram we get a picture like figure \ref{biaenevol}. Notice that all the 
three different cases start off at past light-like infinity 
$\mathcal{I}^-$. They all start off as a Kasner universe (straight line). 
They end up however, at different places. For a positive-valued 
cosmological constant the universe ends as a deSitter universe: moving 
along the $t$-axis to the time-like future $i^+$. The Kasner universe 
($\Lambda=0$) ends at the future light-like infinity $\mathcal{I}^+$, 
while the case $\Lambda<0$ the universe collapses again in past 
light-like infinity $\mathcal{I}^-$. The whole boundary \emph{except} 
past time-like infinity $i^-$ represents singular space-time geometries, 
and thus according to the tunneling boundary condition, is not allowed. 
However, from the calculations done in the previous chapters, there may 
be indications that the starting point $\mathcal{I}^-$ is somewhat 
unstable concerning inclusion of a scalar field. Even a rather modestly 
behaved scalar field effectively reduces the speed of the universe near 
the initial singularity to below ``the speed of light''. The reduction of 
the speed of a Kasner universe is shown in figure \ref{timelikekasner}. 
This effectively changes the path's end-points, so that the universe 
starts off at $i^-$. It ought to be emphasized that this is not enough to 
ensure  regularity of the universe at the starting point, but this may 
indicate that there exists a homotopy of time-like paths connecting the 
Kasner universe with the deSitter solution. These different solutions 
will be universes with different speeds in superspace. One might say that 
we have given the Kasner universe a mass, and the homotopy can be defined 
by the action of the 3-dimensional Lorentz group. 

In the light of this discussion we change the formulation of the 
tunneling proposal \emph{to include all universes which may be connected 
to universes which satisfy the regularity condition, by the action of the 
Lorentz group in superspace.}

\subsection{The tunneling wave function}
The tunneling wave function will have a current in superspace which 
points outwards at the singular boundaries of superspace. The conserved 
current
\[ J=-\frac{i}{2}(\Psi^*\nabla\Psi-\Psi\nabla\Psi^*) \]
which is analogous to the conserved Klein-Gordon current, states that its 
zero- component $J^0$ is a conserved entity. The zeroth component of the 
Klein-Gordon current can be interpreted as a charge, which locally can 
obtain both negative and positive values. Also the zeroth component 
superspace current can be both negative and positive. In light of the 
tunneling proposal, these correspond to expanding and contracting solutions. 

Using the solutions to the WD-equation for a zero-mass scalar field, we 
get the ``space'' components of the superspace current to be:
\begin{equation}
J^{\pm}=-\frac{i}{2}v^{-\frac{2\xi}{3}}(\Psi^*\partial_{\pm}\Psi
-\Psi\partial_{\pm}\Psi^*)=v^{-\frac{2\xi}{3}}k_{\pm}|\Psi|^2
\end{equation}
At the singular boundary $v\rightarrow \infty$, we use the solutions for 
a positive cosmological constant in the big geometry limit. The two 
possible solutions are the Hankel functions $H_p^{(1)/(2)}(v)\approx 
\frac{1}{\sqrt{v}}e^{\pm iv} $.
The zero-component then becomes:
\begin{equation}
J^0\approx \mp |\lambda|^{\frac{1}{2}} v^{2(1-\frac{\xi}{3})}|\Psi|^2
\end{equation}
The outgoing modes are therefore the second Hankel function 
$H_p^{(2)}(v)$ and the tunneling wave functions are are a linear 
combinations of the functions
\begin{equation}\label{tunnel}
\Psi(v, \beta_{\pm}, \Phi)=v^{\frac{\xi}{6}}H_p^{(2)}
(\lambda^{\frac{1}{2}}v)e^{i(k_+\beta_++k_-\beta_-)}e^{in\Phi}
\end{equation}
The solution space of the WD equation will therefore be spanned by the 
solutions (\ref{tunnel}) multiplied by the solutions from the 
$SO(3)$-sector $D^{(l)}_{mm'}$. 

\subsection{Does the Tunneling boundary proposal predict an isotropic 
universe?}
As we have constructed the Tunneling wave function we might wonder if the 
wave function predict an isotropic or an anisotropic universe. In the 
classical case we had trouble with reconstructing the FRW universe in the 
limit $A\longrightarrow 0$. The FRW universe appeared as if it was 
disconnected from the Bianchi Type $I$ solutions. This picture is 
drastically improved in the Quantum case. If we hold $v$ constant, that 
is keeping the volume of the universe constant, and let $A$ approach zero 
from a classical point of view we have to let the function $V(t)$ approach
 infinity. For a positive cosmological constant the limit 
$V\longrightarrow \infty$ is indeed a isotropic FRW limit with finite 
spatial volume. As $v$ and $A$ are independent variables we could 
interpret the $A\longmapsto 0$ a FRW limit. Inserting a scalar field we 
saw that it also reduced the anisotropy. The anisotropy was reduced as the
 ratio $\frac{k^2}{k^2+n^2}$. These two mechanisms for reducing the 
anisotropy affect the classical solutions in two different ways. Whereas 
the reduction of $A$ isotropize the universe through the modular and 
evolutionary degrees of freedom, the inclusion of a scalar field reduces 
the Kasner circle. Thus a scalar field could also work for isotropy 
reduction for the Kasner universe (with infinite spatial sections). 
$A\longrightarrow 0$ for the Kasner universe is however more subtle 
because apparently we do not have any clear {\it geometric} meaning of the
 entity $A$ in that case. 

The $k^2$ enters the wave functions through the order of the Hankel 
function. The order $p$ is given by $p^2=\frac{\xi^2}{36}-k^2-n^2$. The 
factor ordering parameter is assumed to be small, so we assume that $p$ 
is purely imaginary: $p=ir$ where $r\geq 0$. A small $r$ means an 
isotropic universe. 
 From the integral representation of the Hankel functions\cite{GR65} we can
 write 
\[ 
H_{ir}^{(2)}(z)=\frac{ie^{-\frac{r\pi}{2}}}{\pi}\int_{-\infty}^{\infty}
e^{-iz\cosh t-irt}dt
\]
Apparently the wave functions are suppressed by an exponential factor as 
$r$ grows. Thus the probability amplitude is exponentially damped in $r$: 
$|\Psi|^2\sim e^{-r\pi}$. We do have to be a bit careful however because 
any linear combination of these wave functions is a solution. We might 
imagine that it had an exponentially growing prefactor in $r$. Let us try 
to estimate a reasonable prefactor by ``normalizing'' the wave functions 
over superspace. Since the variable $v$ will behave as a time variable in 
superspace we could also say that we divide by the ``time average'' of the
 particular solution. The ``normalizing'' integrals $\int dv |\Psi(v)|^2$ 
 diverge for the Hankel functions. We will therefore do the following: we 
Wick-rotate the volume element, $v\longmapsto iv$ so that the Hankel 
functions $H_p^{(2)}$ are ``rotated'' to the Bessel functions $K_p$. This 
function will under reasonable conditions be square integrable. We 
estimate the prefactor by demanding that the wave function is normalized 
after a Wick-rotation of the volume element. 
The Bessel functions are related through the following equations
\begin{eqnarray}
K_p(z)&=&\frac{\pi}{2}ie^{\frac{\pi}{2}pi}H_p^{(1)}(iz)\\
\overline{H_p^{(2)}}(z)&=&H^{(1)}_p(\overline{z})
\end{eqnarray}
We also introduce the integration measure $\sqrt{|G|}$ which reflects the 
geometry of minisuperspace.  The DeWitt metric (eq. \ref{dewittmetric}) 
will yield $\sqrt{|G|}=v^{\frac{1}{2}}$ but if we assume\footnote{Thus 
these two integration measures coincide iff $\xi=\frac{3}{2}$. We would 
also mention that the integral is manageable for any power of $v$ 
(provided that the integral converges).} $\sqrt{|G|}=v^{1-\frac{\xi}{3}}$ 
the resulting integral become surprisingly simple\cite{GR65}:
\begin{eqnarray}
I&=&\int_0^{i\infty}dv\sqrt{|G|}|\Psi|^2=\frac{4}{\pi^2}e^{\pi r}
\int_0^{\infty}dv v\left|K_{ir}(\lambda^{\frac{1}{2}}v)\right|^2\\
&=& \frac{1}{\lambda\pi^2}\frac{\pi r}{1-e^{-2\pi r}}
\end{eqnarray}
Using these wave functions after ``normalization'' the probability 
amplitude decays as
\begin{eqnarray}
|\Psi|^2\approx \frac{e^{-\pi r}}{2\pi r}\left(1-e^{-2\pi r}\right)
\end{eqnarray}
These wave functions will peak for $r=0$ so they predict an isotropic 
universe. They will exponentially decay for larger values of $r$. Whether 
this ``normalizing'' function is the right weight function is so far only 
speculation, but if these functions are reasonable the wave functions 
predict an isotropic universe. 

\section{Conclusion and Summary}
In this paper we have discussed the Bianchi Type $I$ minisuperspace model 
with compact spatial sections. In the classical case this removed some 
pathologies. The classical solutions were all written down in a way so 
that the solution space was manifestly factorized into three parts: a 
topological sector, a Kasner sector and a matter sector. The Kasner 
solutions were obtained in the limit $A\longrightarrow \infty$ where one 
easily could see that the modular sector becomes degenerate. Choosing 
compact spatial sections also gave a better understanding of the topology 
of the 4-space. For instance the apparently flat case 
$\gamma=0=T^{\mu\nu}$ of the Kasner universe was shown to have a conical 
singularity at $t=0$. 

In the quantum mechanical case since the (true) phase space has a 
dimension equal to the dimension of the solution space, it is easier to 
give a precise meaning to the WD equation. This paper has also shown the 
necessity of introducing compact spatial sections when one wants to 
quantize the model, and provides a solution to the WD equation for $T^3$ 
spatial sections. Also a set of solutions in the $SO(3)$ sector have been 
given which had particularly nice algebraic properties. These solutions
 have a structure which describes a shell structure in the $SO(3)$ 
sector. It was also indicated how these solutions allow geometry changing 
solutions. 

Inserting various matter configurations  in general showed isotropization 
of the universe. The resulting tunneling wave functions appeared as if 
they peaked at small anisotropy. Under reasonable considerations we 
constructed a wave function which had an exponentially decaying 
probability 
amplitude for increasing anisotropy. This seems to agree with other 
authors\cite{huang97,huang98}. 

It is quite clear that universe models with compact spatial sections will 
yield many interesting and surprising results. Since these possibilities  
only recently have come to physicist's and cosmologist's attention there 
is a \emph{lot} of work still to be done. What we have shown in this 
paper is that even at the classical level, this may reveal interesting 
properties of cosmological models. 

\section*{Acknowledgments}
First of all I would like to thank \O. Gr\o n for great inspiration and 
for encouraging me to write this paper. I am also very grateful to \O. 
Gr\o n and B. Steffensen  for reading through the manuscript and making 
useful comments. 
Helpful conversations with  J.M. Leinaas are also gratefully acknowledged.

\end{document}